\documentclass[prb,aps,preprint,double-spaced,eclepsf,amsmath,amssymb,showpacs]{revtex4}  

\usepackage[dvips]{graphicx}
\usepackage{dcolumn}
\usepackage{bm}

\begin{document}                
\draft
\title{Evolution of the electronic structure from the insulator to the superconductor in Bi$_2$Sr$_{2-x}$La$_x$(Ca,Y)Cu$_2$O$_{8+\delta}$}

\author{K.~Tanaka,$^1$ T.~Yoshida,$^1$ K.M.~Shen,$^{2,3}$ D.H.~Lu,$^2$ W.S.~Lee,$^2$ H.~Yagi,$^1$ A.~Fujimori,$^1$ Z.-X.~Shen,$^2$
Risdiana,$^4$ T.~Fujii,$^{4,5}$ and I.~Terasaki$^4$}

\address{$^1$Department of Physics, University of Tokyo, Hongo, Tokyo 113-0033, Japan\\ $^2$Department of  Applied Physics and Stanford
Synchrotron Radiation Laboratory, Stanford University, Stanford, CA
94305, USA\\ $^3$Department of Physics, Laboratory of Atomic and
Solid State Physics, Cornell University, Ithaca NY 14853, USA\\
$^4$Department of Applied Physics, Waseda University, Tokyo
169-8555, Japan\\ $^5$Cryogenic Center, University of Tokyo,
Bunkyo-ku, Tokyo 113-0032, Japan}

\date{\today}

\begin{abstract}                
La-doped and Y-doped Bi$_2$Sr$_2$CaCu$_2$O$_{8+\delta}$ (Bi2212)
compounds Bi$_2$Sr$_{2-x}$La$_x$(Ca,Y)Cu$_2$O$_{8+\delta}$, which
range from the insulator to the deeply underdoped superconductor,
have been studied by angle-resolved photoemission spectroscopy. We
have observed that the lower Hubbard band (LHB) of the parent
insulator is gradually shifted upward with doping without
significantly changing the band dispersions, which implies a
downward shift of the chemical potential with hole doping. This
behaviour is analogous to Bi$_2$Sr$_{2-x}$La$_x$CuO$_{6+\delta}$
(Bi2201) and Ca$_{2-x}$Na$_x$CuO$_2$Cl$_2$ (Na-CCOC) but is
different from La$_{2-x}$Sr$_x$CuO$_4$ (LSCO), where the LHB stays
well below the chemical potential and does not move while its
intensity quickly diminishes in the underdoped region.

\end{abstract}
\pacs{74.72.Hs, 79.60.-i, 71.30.+h} \maketitle

\section{INTRODUCTION}
The question of how the electronic structure of high-$T_c$ cuprates
evolves from the Mott insulator to the superconductor with hole
doping is one of the most fundamental and important issues in
condensed matter physics. This subject has been extensively
investigated experimentally and theoretically, but still remains
highly controversial. Previous angle-resolved photoemission
spectroscopy (ARPES) studies have revealed two different cases. In
underdoped La$_{2-x}$Sr$_x$CuO$_4$ (LSCO), a ``two-component"
electronic structure has been observed, that is, upon hole doping
``in-gap" states appears primarily well ($\sim$0.4 eV) above the
lower Hubbard band (LHB), the chemical potential does not shift and
spectral weight is transferred from the LHB to the in-gap states for
further hole doping within the underdoped
region~\cite{Ino_twocomponent,Yoshida_lightly,Ino_chemipote}.
Already in the lightly-doped region, a weak quasi-particle (QP) peak
crosses the Fermi level ($E_{\rm{F}}$) in the (0,0)-($\pi,\pi$)
nodal direction and is responsible for the metallic
transport~\cite{Yoshida_lightly}. ARPES spectra of underdoped
Ca$_{2-x}$Na$_x$CuO$_2$Cl$_2$ (Na-CCOC)~\cite{Ronning_onecomponent}
and Bi$_2$Sr$_{2-x}$La$_x$CuO$_{6+\delta}$
(Bi2201)~\cite{Hashimoto}, on the other hand, show that upon hole
doping the chemical potential moves to the top of the LHB and
continues to shift downward for further hole doping.

As for the Bi$_2$Sr$_2$CaCu$_2$O$_{8+\delta}$ (Bi2212) family, the
apparently smooth evolution of ARPES spectra in the ($\pi,0$) region
in a combined plot of undoped Ca$_2$CuO$_2$Cl$_2$ (CCOC) and
underdoped Bi2212 implies that a behaviour similar to Na-CCOC is
expected for Bi2212~\cite{Ronning}. A core-level photoemission study
of Bi2212 has indeed shown that the chemical potential is shifted
with doping in underdoped samples~\cite{Harima} unlike LSCO.
Recently, a momentum distribution curve (MDC) analysis of
Bi2201~\cite{Hashimoto} and underdoped Na-CCOC~\cite{Kyle} have
revealed the existence of an additional QP band just above the LHB
crossing $E_{\rm{F}}$. Here, strong electron-phonon interaction has
been proposed to result in the dressed coherent QP band accompanied
by a high-energy incoherent (Frank-Condon type) feature. Yet, the
different behavior of the chemical potential shift and the LHB-QP
energy separation between LSCO and Na-CCOC or Bi2201 remains to be
explained. In order to see what makes those differences between the
different families in the evolution of the electronic structure from
the Mott insulator to the superconductor, we have performed detailed
APRPES measurements of lightly-doped Bi2212.

\section{EXPERIMENTS}
Recently, high-quality crystals of lightly-doped Bi2212 have become
available~\cite{sample_Fujii}, which enabled us to study the
systematic doping dependence covering from the insulator to the
superconductor. In particular, substituting La for the Sr site
instead of substituting for the Ca site has made lightly-doped
Bi2212 samples metallic ($d\rho/dT$$>$$0$) in an analogous way to
the lightly-doped LSCO~\cite{Ando_metallic} and
YBa$_2$Cu$_3$O$_{7-\delta}$(YBCO)~\cite{Ando}. Single crystals of
Bi$_2$Sr$_{2-x}$La$_x$CaCu$_2$O$_{8+\delta}$ and
Bi$_2$Sr$_2$Ca$_{0.8}$Y$_{0.2}$Cu$_2$O$_{8+\delta}$ were grown by
the traveling solvent floating-zone method. X-ray diffraction showed
no trace of impurity phases. Details of the sample preparation are
given elsewhere~\cite{sample_Fujii}. The hole concentration $p$ per
Cu atom was determined using the empirical relationship between $p$
and the room-temperature thermopower~\cite{thermopower}. All the
samples show metallic transport at 300 K, while some of them show
insulating behavior ($d\rho/dT$$<$$0$) at low temperatures. $p$,
$T_{c}$, $T_{\textrm{min}}$ and $\rho_{\textrm{min}}$ of the
measured samples are listed in Table.~\ref{table1}. Here,
$T_{\textrm{min}}$ is the temperature at which the resistivity
reaches the minimum value $\rho_{\textrm{min}}$. The
Bi$_2$Sr$_2$Ca$_{0.8}$Y$_{0.2}$Cu$_2$O$_{8+\delta}$ ($p$$=$$0.075$)
sample was superconducting and above $T_{c}$ $d\rho/dT$$>$$0$.

ARPES measurements were carried out at beamline 5-4 of Stanford
Synchrotron Radiation Laboratory (SSRL). Incident photons had an
energy of ${\it h}{\nu}$$=$$19$ eV. A SCIENTA SES-200 analyzer was
used in the angle mode with the total energy and momentum
resolution of $\sim$14 meV and $\sim$$0.3^\circ$, respectively.
Samples were cleaved \textit{in situ} under an ultrahigh vacuum of
$10^{-11}$ Torr, and measured at $\sim$$10$ K. The position of the
Fermi level was calibrated with gold spectra.

\section{RESULTS AND DISCUSSION}
Figure~1(a)-(d) shows energy distribution curves (EDCs) along the
diagonal $\textit{\textbf{k}}$=$(0,0)$-$(\pi,\pi)$ direction (nodal
direction) in the second Brillouin zone (BZ).  The intensity maps in
$E$-$k$ space shown in Figs~1(e)-(h) reveal the peak dispersion.
First, the spectra for $p$=$0.03$ show a single dispersive peak
marked by ``\textit{MB} (main band)" plus a diffraction replica
marked by ``\textit{SS}" due to the Bi-O plane superstructure. This
peak disperses closest to $E_{\rm{F}}$ around $\sim$($\pi/2,\pi/2$),
and can be considered as a remnant of the LHB or more precisely of
the Zhang-Rice singlet band. With hole doping $p$, the dispersive
peak as a whole moves upward until an obvious $E_{\rm{F}}$ crossing
in the nodal direction occurs for $p$$\sim$$0.075$, where the system
becomes superconducting. This is contrasted with spectra of LSCO
with similar doping levels, where the LHB stays away ($\sim$$-0.5$
eV) from $E_{\rm{F}}$ but a sharp QP peak crossing $E_{\rm{F}}$ is
visible already for $p$$\simeq$$0.03$~\cite{Yoshida_lightly} due to
the presence of two separate spectral features, namely, the LHB and
the QP crossing $E_{\rm{F}}$.

Figure~2(a)-(d) shows the ARPES spectra along the ``underlying"
Fermi surface~\cite{Ronning}. Here, by the ``underlying" Fermi
surface is meant the minimum-gap locus including the pseudogap
region around $\textit{\textbf{k}}$$\simeq$$(\pi,0)$ as reported for
underdoped Bi2212~\cite{Ding}. The figure again shows a single
dispersive feature that moves upward with hole doping $p$. This is
also contrasted with the case of LSCO where the LHB stays
$\sim$$0.5$ eV below $E_{\rm{F}}$, where the QP crosses, leading to
the ``two-component" behavior, in particular around
($\pi,0$)~\cite{Ando}. While a dispersive peak which crosses
$E_{\rm{F}}$ occurs in the $(0,0)$-$(\pi,\pi)$ nodal direction (for
$p$$\ge$0.075), no $E_{\rm{F}}$ crossing occurs and a finite
(pseudo)gap persists around $\textit{\textbf{k}}$$\sim$($\pi,0$).
The second derivatives of the spectra shown in Fig.~2(e)-(h)
indicate the dispersion along the underlying Fermi surface.

In Fig.~3, we have plotted the doping dependence of the peak
positions in EDCs on the (underlying) Fermi surface at
$\sim$($\pi/2,\pi/2$) and those at $\sim$($\pi,0$) thus determined
by taking the second derivatives. One can see that in Bi2212, the
peak positions at $\sim$($\pi/2,\pi/2$) and $\sim$($\pi,0$) show
nearly parallel shifts with doping in the lightly-doped region,
again indicating a rigid-band-like shift of the LHB similar to the
case of Na-CCOC~\cite{Kyle} and Bi2201~\cite{Hashimoto}. In the same
figure, we have plotted the doping dependence of the peak position
at $\sim$($\pi/2,\pi/2$) and $\sim$($\pi,0$) in LSCO determined in
the same way~\cite{Yoshida}. In LSCO a sharp QP feature appears near
$E_{\rm{F}}$ already at $p$$\sim$$0.03$ and stays there while the
LHB is located at $\sim$$-0.5$ eV. The amount of the shift of each
feature is much smaller in LSCO than that in Bi2212 and Na-CCOC.

To reveal further details of the evolution of the electronic
structure near $E_{\rm{F}}$, EDCs along $(0,0)$-$(\pi,\pi)$ of
lightly-doped Bi2212 are plotted on an expanded scale in
Fig.~4(a)-(d) with the peak positions determined by the second
derivatives of the EDCs. Figure 4(e) shows the dispersion of the EDC
peak (representing that of the LHB) along the (0,0)-($\pi,\pi$)
direction. In addition, we have plotted the dispersion of the ``QP
band" which has been determined by the momentum distribution curve
(MDC) analysis, as performed for Bi2201~\cite{Hashimoto} and
Na-CCOC~\cite{Kyle}. Although there is no sharp peak crossing
$E_{\rm{F}}$ in the EDCs of the $p$=0.03 and 0.05 samples, the peak
in the MDCs shows a clear dispersion crossing $E_{\rm{F}}$, as shown
in Fig. 4(e). Here, the results for each composition have been
shifted so that the LHB positions coincide. One can see that the
chemical potential thus obtained is shifted downward relative to the
LHB with hole doping. One can also see that the valence-band maximum
(VBM) of the LHB is shifted in $k$-space from $\sim$($\pi/2,\pi/2$)
toward (0,0) with hole doping. Remarkably, the QP band stays almost
at the same position in the $E$-$k$ space in this plot and the
$E_{\rm{F}}$ crossing point, namely, $k_{\rm{F}}$ in the nodal
direction is shifted toward (0,0) following the downward chemical
potential with hole doping. From Figs.~1-4 one can conclude that the
lightly-doped Bi2212 shows a rigid-band-like shift of the LHB
(except for the small shift toward
$\textit{\textbf{k}}$$\sim$($0,0$)) and the nodal QP just above the
LHB, analogous to the case of Bi2201 and Na-CCOC. This behavior is
different from LSCO, where the LHB stays almost at the same energy
and the clear QP band is seen well ($\sim$$-0.5$ eV) above the LHB.

In the antiferromagnetic (AF) insulating state of the undoped
compound, the maximum of the LHB along the nodal direction should
occur on the AF zone boundary, that is, exactly at ($\pi/2,\pi/2$)
due to the folding of the BZ in the AF state. Without the long-range
order, on the other hand, $k_{\rm{F}}$ can in principle take any
value. Therefore, it is interesting to see how the $k_{\rm{F}}$
evolves as a function of $p$. In order to see this, we have plotted
in Fig.~4(f) the doping dependence of the $k_{\rm{F}}$ position
determined from the MDC peak position at $E_{\rm{F}}$ as a function
of $p$. One can see from the figure that the LHB in lightly-doped
Bi2212 indeed moves toward ($\pi/2,\pi/2$) with decreasing hole
concentration. The results for Na-CCOC ($p$=0, 0.05, 0.1, 0.12) also
showed the same behavior~\cite{Kyle}. On the other hand,
$k_{\rm{F}}$ in LSCO extrapolates to $\sim$$(0.44\pi, 0.44\pi)$ and
not to ($\pi/2,\pi/2$) until $p$$\simeq$0.03. (For undoped
La$_2$CuO$_4$, a tiny amount of holes are doped due to excess
oxygens and faint $E_{\rm{F}}$ spectral weight appears at
($\pi/2,\pi/2$)). This can be understood as due to the separated
features of the LHB and the QP band crossing $E_{\rm{F}}$ in LSCO.
This observation again points to the similarity between Bi2212,
Bi2201 and Na-CCOC concerning the evolution of the electronic
structure.

From the present data, one can also obtain an important implication
for the doping dependence of spectra across the
insulator-superconductor transition (between $p$=0.06 and 0.075) in
Bi2212. Nodal spectra in the superconducting phase ($p$=0.075) have
a sharp QP peak around $E_{\rm{F}}$ without a gap, whereas nodal
spectra in the insulating phase ($p$=0.06) have finite shoulder-like
spectral weight (weak and broad QP peak) around $E_{\rm{F}}$ with a
$\sim$$9$ meV leading-edge gap. It therefore appears that the system
shows superconductivity when the nodal spectra show a sharp QP peak
crossing $E_{\rm{F}}$. This behavior is consistent with the recently
proposed two-gap scenario~\cite{twogap,Hashimoto_LSCO} that
superconductivity occurs in the nodal region in underdoped samples.
Our observation indeed implies a close relationship between the
occurrence of superconductivity and the existence of QP peak around
the node. Here, it should be noted again that one can observe a
nodal QP peak already for $p$$\simeq$$0.03$ in the insulating LSCO.
Thus, it seems that the doping evolution of the electronic structure
is whether a QP feature appears near $E_{\rm{F}}$ well above the LHB
before a significant chemical potential shift occurs or not. So far,
LSCO has been the only example that shows the appearance of a QP
band before the chemical potential starts to shift in the
lightly-doped region. It is an important question to ask which makes
this difference between the different families of cuprates. It has
been pointed out that the major difference in the electronic
structure of the CuO$_2$ plane between LSCO and Bi2212 lies in the
magnitude of the next-nearest-neighbor hopping $t^{\prime}$ within
the single band description of the CuO$_2$
plane~\cite{t'_BSCCO,t'_Pavarini}. The magnitude of $t^{\prime}$ is
expected to decrease with influence of the apical oxygen on the
CuO$_2$ plane~\cite{Andersen}, and therefore LSCO, which has the
shortest Cu-apical oxygen distance among these cuprate families, is
expected to have the smallest $|t^{\prime}|$. According to the
$t$-$t^{\prime}$-$t^{\prime\prime}$-$J$ model calculation, a larger
energy shift of the chemical potential with hole doping in the
underdoped region has been predicted for larger
$|t^{\prime}|$~\cite{Tohyama}, consistent with our observation that
Bi2212 and Na-CCOC showed faster chemical potential shift than LSCO.

Recently, the new interpretation of the ARPES line shapes of undoped
and lightly-doped cuprates in Na-CCOC and LSCO was
proposed~\cite{Kyle,Gunnarsson,Mishchenko}, which took into account
polaronic effects because arising from strong electron-phonon
coupling. The broad hump structure was successfully explained by polaronic effects.
The similarity of the spectral lineshape between
lightly-doped Bi2212 and Na-CCOC indicates that the polaronic
scenario can also be applied to Bi2212, too. Moreover, theoretical
works reported that the phase separation, which occurs with suitable
electron-phonon coupling strength~\cite{Capone}, successfully
explained the doping evolution of the electronic structure of
underdoped LSCO, namely, the ``two-component"
behavior~\cite{phase_separation}. The calculation also predicted
that phase separation would be suppressed when the electron-phonon
coupling becomes stronger and the system shows polaronic insulating
state~\cite{Capone}. This implies that Bi2212, Bi2201 and Na-CCOC
have larger electron-phonon interaction than LSCO and the different
strength of electron-phonon coupling makes different doping
evolution of the electronic structure in the different families of
cuprates.

\section{SUMMARY}
We have observed the evolution of the electronic structure with hole
doping in lightly-doped Bi2212 from the insulator (with
high-temperature metallic behavior) to the superconductor. The
results show rigid-band-like shifts of (the remnant of) the LHB with
hole doping. The chemical potential is shifted downward and a QP
feature appears around $E_{\rm{F}}$ just above the LHB. This
evolution of the electronic structure, together with the shift of
the momentum position of the maximum of the LHB, are similar to
those reported for Bi2201 and Na-CCOC but are different from LSCO.
In order to establish whether the different $t^{\prime}$'s and
different strength of electron-phonon coupling are responsible for
the different doping evolution of LSCO and Bi2212, systematic
studies on other cuprate families as well as further theoretical
studies are desired.

\section{ACKNOWLEDGMENTS}
We acknowledge technical help by N.\ P.\ Armitage. This work was
supported by a grant-in-Aid for Scientific Research in Priority Area
``Invention of Anomalous Quantum Materials" (16076208) from MEXT,
Japan. SSRL is operated by the DOE Office of Basic Energy Science
Divisions of Chemical Sciences and Material Sciences.

\newpage

\begin{table}[tb]
\caption{Chemical compositions, hole concentration $p$, $T_c$,
$T_{\textrm{min}}$ and $\rho_{\textrm{min}}$ of Bi2212 samples
studied in the present work.} \label{table1}
\begin{center}
\begin{ruledtabular}
\begin{tabular}{ccccc}
Bi$_2$Sr$_{2-x}$La$_x$CaCu$_2$O$_{8+\delta}$&$p$&$T_c$(K)&$T_{\textrm{min}}$(K)&$\rho_{\textrm{min}}$(m$\Omega$cm)\\
\hline
$x=0.8$ & 0.03 & - & 150& 5\\
$x=0.7$ & 0.05 & - & 130& 3.2\\
$x=0.6$ & 0.06 & - & -& -\\
\hline
Bi$_2$Sr$_2$Ca$_{0.8}$Y$_{0.2}$Cu$_2$O$_{8+\delta}$ & 0.075 & 30& - & -\\
\end{tabular}
\end{ruledtabular}
\end{center}
\end{table}

\newpage

\begin{figure}[tb]
\begin{center}
\includegraphics[width=140mm]{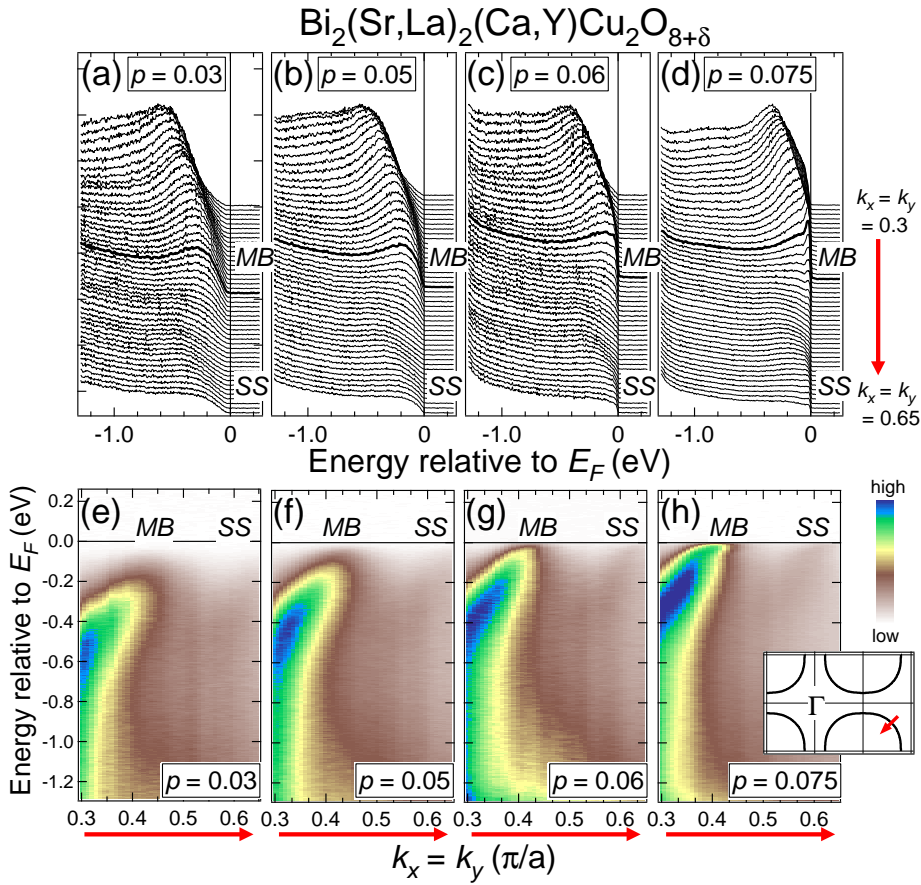}
\caption{ARPES spectra of lightly-doped Bi2212 along the
(0,0)-($\pi,\pi$) nodal direction in the second BZ. (a)-(d): EDCs.
A bold line for each doping indicates the spectrum where the
dispersive feature comes closest to $E_{\rm{F}}$. (e)-(h):
Intensity plot of the spectra in the $E$-$k$ plane. The features
denoted by \textit{SS} are diffraction replica of the main band
(\textit{MB}) due to the Bi-O plane superstructure.}
\end{center}
\end{figure}

\newpage

\begin{figure}[thb]
\begin{center}
\includegraphics[width=130mm]{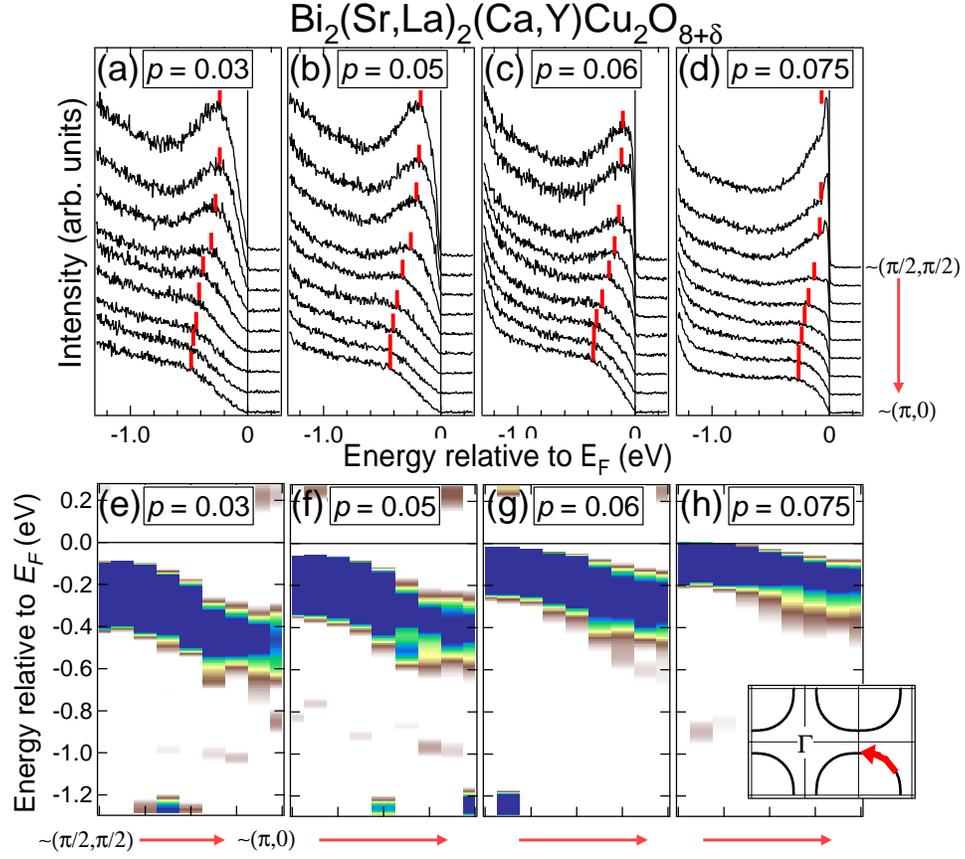}
\caption{ARPES spectra of lightly-doped Bi2212 along the
``underlying" Fermi surface in the second BZ. (a)-(d): EDCs. The
peak position determined by the second derivatives are shown by
vertical bars. (e)-(h): Second derivatives of the EDCs in the
$E$-$k$ plane.}
\end{center}
\end{figure}

\newpage

\begin{figure}[tb]
\begin{center}
\includegraphics[width=100mm]{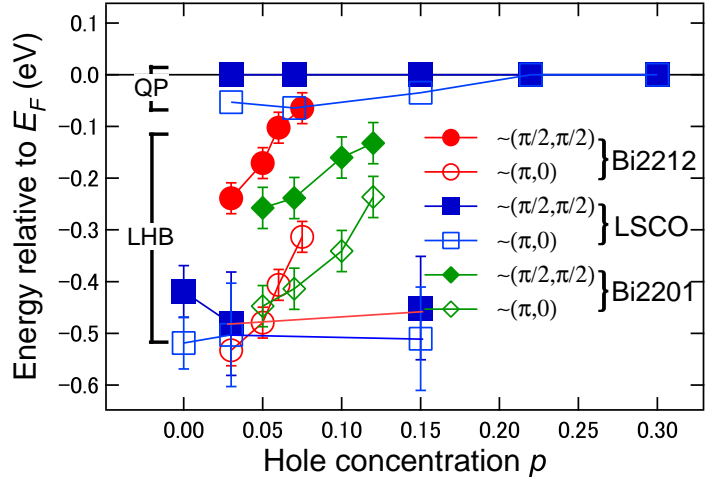}
\caption{Energy positions of the EDC peaks at
$\sim$($\pi/2,\pi/2$) and $\sim$($\pi,0$) in Bi2212,
LSCO~\cite{Yoshida} and Bi2201~\cite{Hashimoto} as functions of
doping levels.}
\end{center}
\end{figure}

\newpage

\begin{figure}[tb]
\begin{center}
\includegraphics[width=120mm]{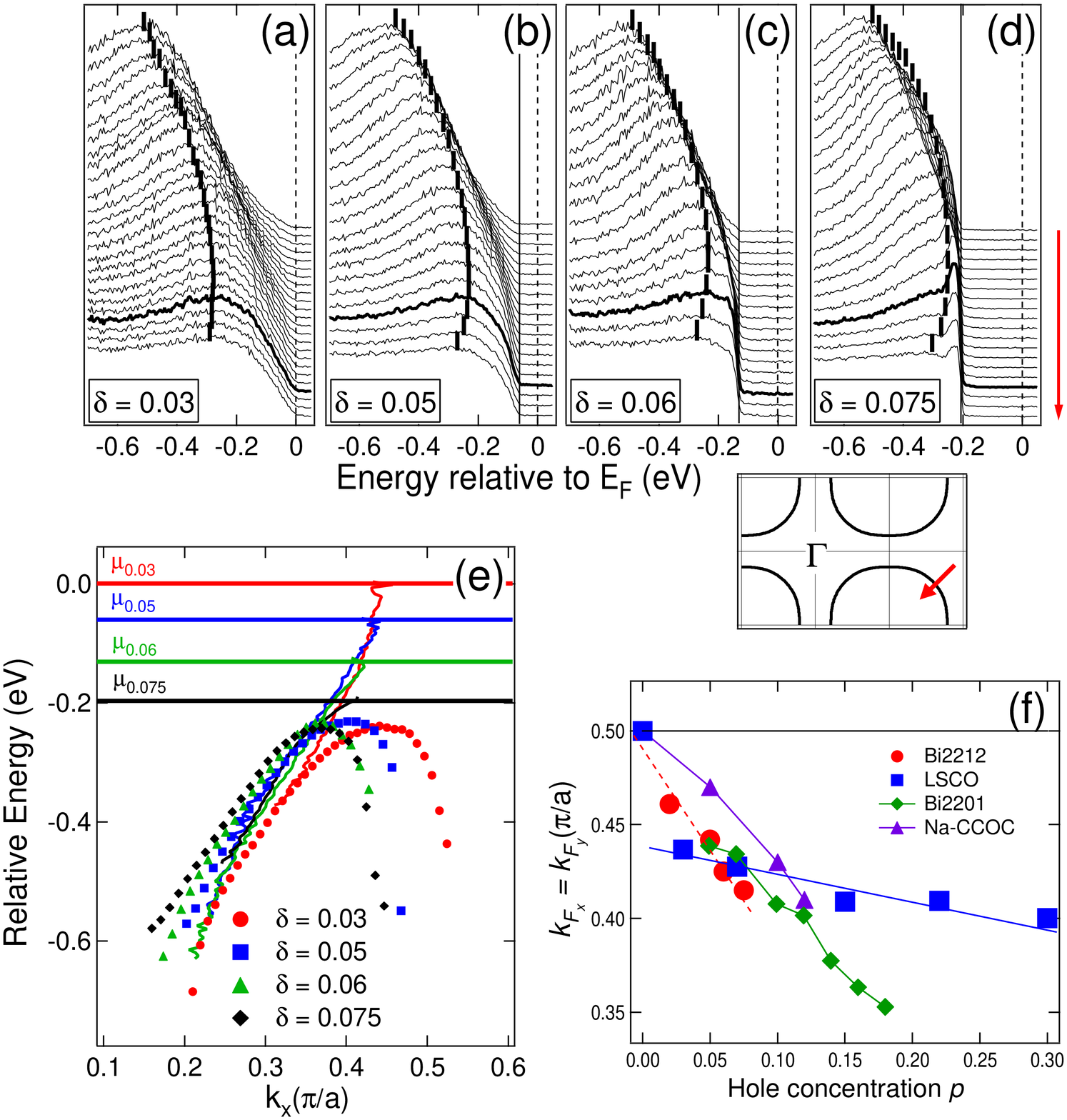}
\caption{Doping evolution of the electronic structure along the
nodal direction near $E_{\rm{F}}$. (a)-(d): Blow up of Fig.~1 near
$E_{\rm{F}}$. Vertical bars indicate the position of the LHB. (e):
Position of the LHB and the ``QP" band obtained from MDCs. The
plots have been shifted vertically so that the chemical potential
$\mu$ shifts the average of the shift in the LHB at
$\sim$($\pi/2,\pi/2$) and $\sim$($\pi,0$) in Fig.~3 for each
doping. The chemical potential $\mu$ is also displayed. (f):
Doping dependence of Fermi momentum $k_{\rm{F}}$ positions in
Bi2212, LSCO~\cite{Yoshida}, Bi2201~\cite{Hashimoto} and
Na-CCOC~\cite{Kyle}.}
\end{center}
\end{figure}

\end{document}